\documentclass[final,3p]{elsarticle}

\usepackage{graphicx}      
\usepackage{natbib}        
\usepackage{algorithm,algpseudocode}
\usepackage{enumerate}
\usepackage{amsmath}
\usepackage{amssymb}
\usepackage{lineno}
\modulolinenumbers[5]

\biboptions{authoryear}

\begin{document}
\begin{frontmatter}

\title{An epidemiological model for the spread of COVID-19: A South African case study}

\author[MOYO,UP]{Laurentz E. Olivier} 
\author[UP]{Ian K. Craig \corref{cor1}} 

\address[MOYO]{Moyo Africa, 
   Centurion, South Africa.}
\address[UP]{University of Pretoria, 
   Pretoria, 0002, South Africa.}   
   
\cortext[cor1]{Corresponding author. Address: Department of Electrical, Electronic, and Computer Engineering, University of Pretoria, Pretoria, South Africa. \\Tel.: +27 12 420 2172; fax: +27 12 362 5000.}

\begin{abstract}                
An epidemiological model is developed for the spread of COVID-19 in South Africa. A variant of the
classical compartmental SEIR model, called the SEIQRDP model, is used. As South Africa is still in the
early phases of the global COVID-19 pandemic with the confirmed infectious cases not having
peaked, the SEIQRDP model is first parameterized on data for Germany, Italy, and South Korea --
countries for which the number of infectious cases are well past their peaks. Good fits are achieved
with reasonable predictions of where the number of COVID-19 confirmed cases, deaths, and
recovered cases will end up and by when. South African data for the period from 23 March to 8 May
2020 is then used to obtain SEIQRDP model parameters. It is found that the model fits the initial
disease progression well, but that the long-term predictive capability of the model is rather poor.
The South African SEIQRDP model is subsequently recalculated with the basic reproduction number
constrained to reported values. The resulting model fits the data well, and long-term predictions
appear to be reasonable. The South African SEIQRDP model predicts that the peak in the number of confirmed infectious individuals will occur at the end of October 2020, and that the total number of deaths will range from about 10,000 to 90,000, with a nominal value of about 22,000. All of these predictions are heavily dependent on the disease control measures in place, and the adherence to these measures. These predictions are further shown to be particularly sensitive to parameters used to determine the basic reproduction number. The future aim is to use a feedback control approach together with the South African SEIQRDP model to determine the epidemiological impact of varying lockdown levels proposed by the South African Government.
\end{abstract}


\end{frontmatter}

\section{Introduction}

A novel coronavirus emerged in Wuhan, China towards the end of 2019. This virus, which was subsequently named SARS-CoV-2 and the disease it causes COVID-19 \citep{WHO1}, has since spread around the world. The WHO characterized COVID-19 as a pandemic on 11 March 2020 \citep{WHO2}. The COVID-19 pandemic first took hold in regions of the world that share high volumes of air traffic with China \citep{Lau:20}. The worst affected countries are the US, Britain, Italy, France, and Spain who have reported the highest number of COVID-19 deaths to date \citep{Economist:20}. The importance of ``flattening the curve'', i.e. reducing the number of COVID-19 infected patients needing critical care to be below the number of available beds in intensive care units, soon became evident \citep{Stewart:20}.

The South African National Institute for Communicable Diseases confirmed the first COVID-19 case in South Africa on 5 March 2020, a 38-year-old male who recently returned from Italy \citep{NICD}. Having learnt from elsewhere about the importance of ``flattening the curve'', the South African Government was quick to place the country under strict lockdown on 27 March 2020 after only 1,170 confirmed COVID-19 cases and 1 related death \citep{HDX:20}. The early strict lockdown measures have been successful in stemming the spread of the disease. By 8 May 2020 the number of confirmed cases in South Africa were 8,895, a compound daily growth rate of about 5\% since the start of the lockdown on 27 March 2020. In contrast, the confirmed COVID-19 cases in Italy went from a similar base of 1,128 cases on 29 February 2020 to 9,172 cases on 9 March 2020, a compound daily growth rate of about 23\%.

The strict lockdown measures in South Africa have been successful from an epidemiological point of view, but great harm has been done to an economy that was already weak before the COVID-19 pandemic started \citep{Arndt:20}. As a result, pressure is building to relax the lockdown measures \citep{BBC}. There is thus significant interest in determining the epidemiological impact of the lockdown levels proposed by the South African Government \citep{LockdownLevels}. Towards this end, an epidemiological model is developed for South Africa in this work.  

Epidemiological models are useful in that they can predict the progression of an epidemic, and enhance understanding of the impact that infectious disease control measures, such as vaccination and quarantining, may have on reducing the level of infection in a population \citep{Keeling:08}. A variant of the classical compartmental epidemiological SEIR model (see e.g. \cite{Hethcote:91}), called the SEIQRDP model, is used here to model the spread of COVID-19 in South Africa. This model was introduced by \cite{Peng:20} specifically to model the COVID-19 pandemic, and was subsequently used to model the pandemic in Iraq \citep{AlHussein:20} and the USA \citep{Xu:20}. The model is parameterized from data available from The Humanitarian Data Exchange, compiled by the Johns Hopkins University Center for Systems Science and Engineering (JHU CCSE). Available data include COVID-19 confirmed cases, deaths, and recovered cases.

South Africa is still in the early phases of the global COVID-19 pandemic compared to many countries in the Northern Hemisphere. Importantly, confirmed infectious cases in South Africa have yet to peak, whereas such cases are well past their peaks in countries such as Germany, Italy, and South Korea. The SEIQRDP model is therefore first parameterized on data for these three countries in order to test the efficacy of the model. Good fits are achieved with reasonable predictions of where the number of COVID-19 confirmed cases, deaths, and recovered cases will end up and by when.

South African data for the period from 23 March to 8 May 2020 is used to obtain SEIQRDP model parameters. It is found that the model fits the initial disease progression well, but that the long-term predictive capability of the model is rather poor. A similar scenario is observed for Germany when only initial data, from 11 March to 28 March 2020, are used to determine where German cases will end up. The South African SEIQRDP model was subsequently recalculated with the basic reproduction number constrained to reported values \citep{Lai:20}. The resulting model still fits the data well, and long-term (into 2021) predictions appeared much more reasonable. The study concludes with a sensitivity analysis that show that the model is particularly sensitive to parameters used to determine the basic reproduction number.

The aim is to in future work use the South African SEIQRDP model to determine the epidemiological impact of the different lockdown levels proposed by the South African Government \citep{LockdownLevels}, using a feedback control approach similar to \cite{Stewart:20}.

\section{SEIQRDP model}
\label{sec:SEIQRDP_model}

\begin{figure}
\begin{center}
\includegraphics[width=10cm]{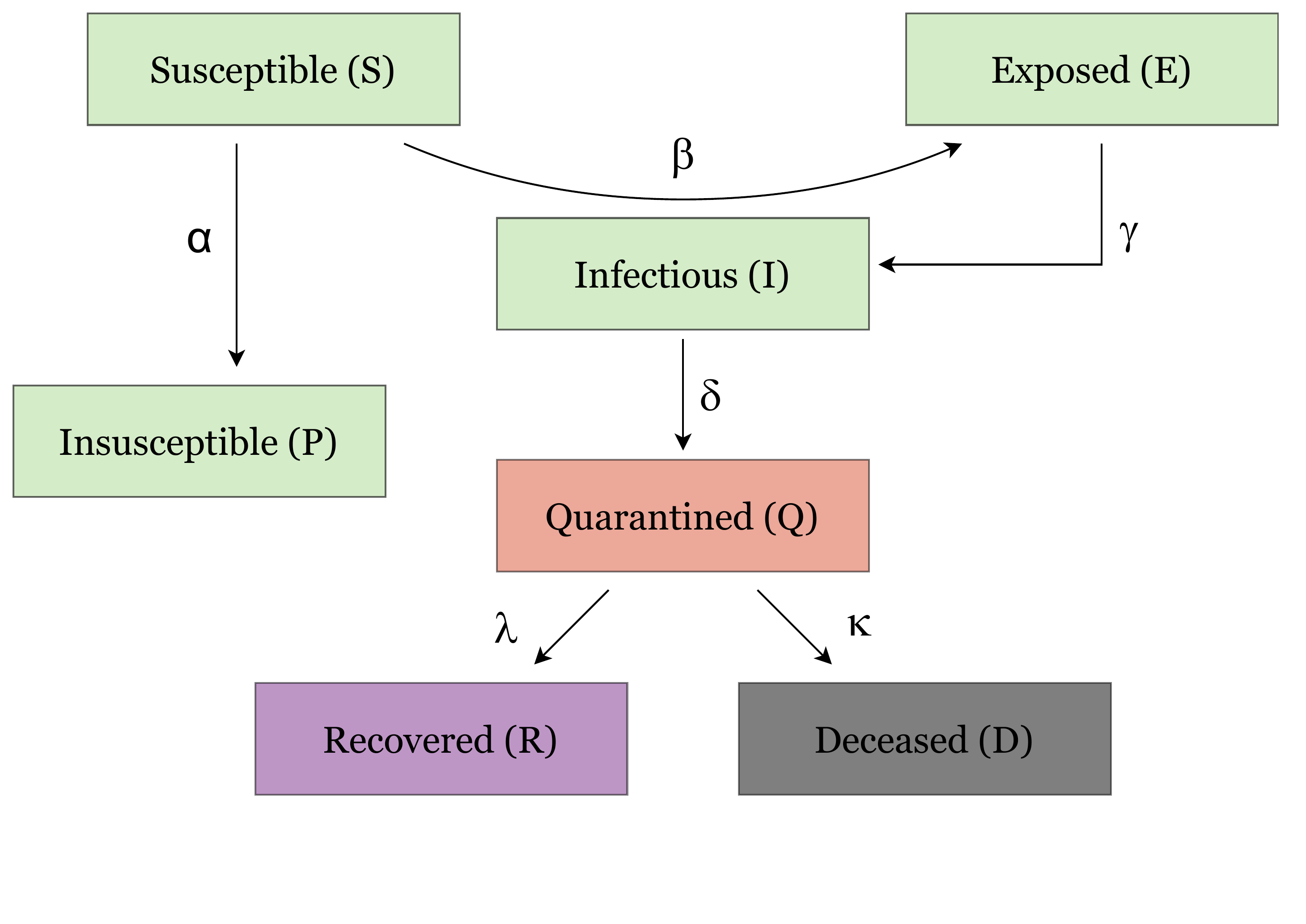}    
\caption{SEIQRDP model.} 
\label{fig:SEIQRDP_model}
\end{center}
\end{figure}

The SEIQRDP model is a generalized compartmental epidemiological model with 7 states. The model was proposed by \cite{Peng:20} and is an adaptation of the classical SEIR model (see e.g. \cite{Hethcote:91}). A numerical implementation of the SEIQRDP model in MATLAB is provided by \cite{Cheynet:20}. The implementation used here is similar to \cite{Cheynet:20}, but model parameter limits and some parameter equations are different as presented in the rest of this paper. The model states and model parameters that drive transitions between them are shown in Fig.~\ref{fig:SEIQRDP_model}. The colours used for Q, R, and D correspond to what is used in the results figures later in the article. The states are described as:
\begin{itemize}
\item S - Portion of the population still susceptible to getting infected,
\item E - Population exposed to the virus; they are infected but not yet infectious,
\item I - Infectious population; infectious but not yet confirmed infected,
\item Q - Population quarantined; confirmed infected,
\item R - Recovered,
\item D - Deceased,
\item P - Insusceptible population.
\end{itemize}

The model equations are given as:

\begin{align}
\label{eq:SEIQRDP_model}
\frac{dS(t)}{dt} & = -\alpha S(t) - \frac{\beta(t)}{N} S(t)I(t) \\
\frac{dE(t)}{dt} & = -\gamma E(t) + \frac{\beta(t)}{N} S(t)I(t) \\
\frac{dI(t)}{dt} & = \gamma E(t) - \delta I(t) \\
\frac{dQ(t)}{dt} & = \delta I(t) - \left( \lambda(t) + \kappa(t) \right) Q(t) \\
\frac{dR(t)}{dt} & = \lambda(t) Q(t) \\
\frac{dD(t)}{dt} & = \kappa(t) Q(t) \\
\frac{dP(t)}{dt} & = \alpha S(t)
\end{align}
where $N$ is the total population size, $\alpha$ is the rate at which the population becomes insusceptible (in general either through vaccinations or medication). At present there is no vaccine that will allow an individual to transfer from the susceptible to insusceptible portion of the population \citep{Promp:20}. Consequently $\alpha$ should be considered to be close to zero. $\beta(t)$ is the (possibly time dependent) transmission rate parameter, $\gamma = 1/N_{lat}$ is the inverse of the average length of the latency period before a person becomes infectious (in days), $\delta = 1/N_{inf}$ is the inverse of the number of days that a person stays infectious without yet being diagnosed, $\lambda(t)$ is the recovery rate, and $\kappa(t)$ is the mortality rate. Both $\lambda(t)$ and $\kappa(t)$ are potentially functions of time, and \cite{Peng:20} notes that $\lambda(t)$ gradually increases with time while $\kappa(t)$ decreases with time. As such, the functions shown in (\ref{eq:lambda}) and (\ref{eq:kappa}) are used to model $\lambda(t)$ and $\kappa(t)$. In (\ref{eq:lambda}) it is set that $\lambda_1 \geq \lambda_2$ such that $\lambda(t) \geq 0$.

\begin{align}
\label{eq:lambda}
\lambda (t) & = \lambda_1 - \lambda_2 \exp(-\lambda_3 t) \\
\kappa (t) & = \kappa_1 \exp(-\kappa_2 t).
\label{eq:kappa}
\end{align}

$\beta$ is often considered to be constant, but is dependent on interventions like social distancing, restrictions on travel, and shutting down of some industries \citep{RSA_Gov:20}. This implies that $\beta$ may also be time dependent. Given that overall restrictions have been made increasingly stringent to curb the spread of the disease, $\beta (t)$ is modelled using a decreasing function of time of the form

\begin{equation}
\beta (t) = \beta_1 + \beta_2 \exp(-\beta_3 t) . \\
\end{equation}

The basic reproduction number, $R_0$, which is the expected number of cases directly generated by one case in the population, is given by \cite{Peng:20} as
\begin{equation}
R_0 = \frac{\beta (t)}{\delta} \left( 1 - \alpha \right) ^ T,
\end{equation}
where $T$ is the number of days. When $\alpha \approx 0$, this can be simplified as
\begin{equation}
\label{eq:R0}
R_0 \approx \frac{\beta (t)}{\delta}.
\end{equation} 

Interventions such as social distancing, restrictions on population movement, wearing of masks (among others) can reduce the effective reproduction number mainly through reducing the effective number of contacts per person. The effective reproduction number found through modelling may therefore likely be lower than what epidemiologists report; e.g. \cite{Lai:20} notes that the reproduction number for COVID-19 is between 2.24 and 3.58.

\section{Parameter estimation}
\label{sec:parameter_estimation}

Data are obtained from The Humanitarian Data Exchange\footnote{Accessible from https://data.humdata.org/dataset/novel-coronavirus-2019-ncov-cases}, as compiled by the Johns Hopkins University Center for Systems Science and Engineering (JHU CCSE) from various sources. The data include the number of confirmed infectious cases, recovered cases, and deceased cases per day from January 2020.

In order to get a sense of the applicability of the model and what the parameter values should be, parameter estimations are first carried out to determine SEIQRDP models for Germany, Italy, and South Korea. These countries are selected as their outbreaks started earlier than that of South Africa, and consequently their parameter estimation should be more accurate. They have also had differing approaches under different circumstances, which means that the different parameters obtained should illustrate how the model behaves.

The parameter values obtained are shown in Table~\ref{tb:fitting_parameters}, along with the allowed low and high limits. The limits are important to ensure that the model not only provides a good fit to the data, but that the behaviour is kept within what is epidemiologically expected.

Given that there is currently no vaccine against COVID-19 \citep{Promp:20}, $\alpha$ should remain very close to zero. \cite{Lai:20} notes that the basic reproduction number ($R_0$) is between 2.24 and 3.58. A dynamic constraint is therefore used to keep $R_0 \approx \frac{\beta(t)}{\delta} < 3.58$.

\cite{Li:20} notes that the mean latency period for COVID-19 is 3.69 days; $\gamma$ is therefore limited to $\gamma \in \left[ 0.2 , 0.5 \right]$ to yield a latency period between 2 and 5 days.

\cite{Lai:20} notes that the mean incubation period for COVID-19 is between 2.1 and 11.1 days. This implies that
\begin{equation}
2.1 < \frac{1}{\gamma} + \frac{1}{\delta} < 11.1
\end{equation}
and therefore that $1 < \frac{1}{\delta} < 9.1$. The limits are set to $\delta \in [0.1 , 1]$ to yield value between 1 and 10 days. 

Given that $\lambda$ and $\kappa$ are daily rates, they are kept between 0 and 1 to ensure a proper fit to the ``recovered'' and ``deceased'' data. \cite{AlHussein:20} and \cite{Peng:20} found the majority of their $\lambda$ and $\kappa$ values to be between 0 and 0.1.

Also included in Table~\ref{tb:fitting_parameters} are two parameters to quantify the goodness-of-fit to the data. Firstly is the mean squared error (MSE) defined by:
\begin{equation}
MSE = \frac{1}{n} \sum_{i=1}^{n} \left\Vert \mathbf{y}_i - f(\mathbf{x}_i) \right\Vert
\end{equation}
where $n$ is the number of data points, $\mathbf{y}_i$ the output data, and $f(\mathbf{x}_i)$ the output of the model evaluated at the parameters $\mathbf{x}$ found during the fitting step. Also included is the coefficient of determination ($R^2$), which indicates the proportion of the variance in the output data predictable using the independent variables.

\begin{table*}[t]
\begin{center}
\caption{Model parameters by region.}
\label{tb:fitting_parameters}
\begin{tabular}{cccccccc}
Param. & Min & Max & Germany & Italy & South Korea & South Africa \\\hline \\[-7pt]
$\alpha$    & 0     & $\mathrm{10^{-6}}$ & $\mathrm{5 \times 10^{-7}}$ & $\mathrm{10^{-6}}$ & $\mathrm{4 \times 10^{-9}}$ & $\mathrm{3 \times 10^{-11}}$ \\
$\beta_1$   & 0     & $R_0 < 3.58$    & 0.037 & 0.343 & 0.196 & 0.250 \\
$\beta_2$   & 0     & $R_0 < 3.58$    & 0.956 & 0.646 & 1.298 & 0.364 \\
$\beta_3$   & 0     & $R_0 < 3.58$    & 0.087 & 0.133 & 0.264 & 0.003 \\ 
$\gamma$    & 0.2   & 1       		  & 0.337 & 0.269 & 0.500 & 0.200 \\
$\delta$    & 0.1   & 1       		  & 0.209 & 0.407 & 0.417 & 0.464 \\
$\lambda_1$ & 0     & 1       		  & 0.084 & 0.040 & 0.095 & 0.039 \\
$\lambda_2$ & 0     & 1       		  & 0.055 & 0.033 & 0.095 & 0.039 \\
$\lambda_3$ & 0     & 1       		  & 0.036 & 0.013 & 0.015 & 0.066 \\
$\kappa_1$  & 0     & 1       		  & 0.002 & 0.011 & 0.001 & 0.002 \\
$\kappa_2$  & 0     & 1       		  & 0.000 & 0.010 & 0.007 & 0.000 \\
$\mathrm{N}$ (million)	& - & - 	  & 82.93 & 60.43 & 51.64 & 57.78 \\[5pt]
\hline \\[-7pt]
MSE & 0 & $\infty$ & $\mathrm{5.73 \times 10^{7}}$ & $\mathrm{1.76 \times 10^{7}}$ & $\mathrm{3.78 \times 10^{5}}$ & $\mathrm{9.18 \times 10^{4}}$ \\
$\mathrm{R^2}$ & $-\infty$ & 1 & 0.981 & 0.992 & 0.979 & 0.965 \\
\hline
\end{tabular}
\end{center}
\end{table*}

\subsection{Parameter estimation for Germany, Italy, and South Korea}

\begin{figure}
\begin{center}
\includegraphics[width=12cm]{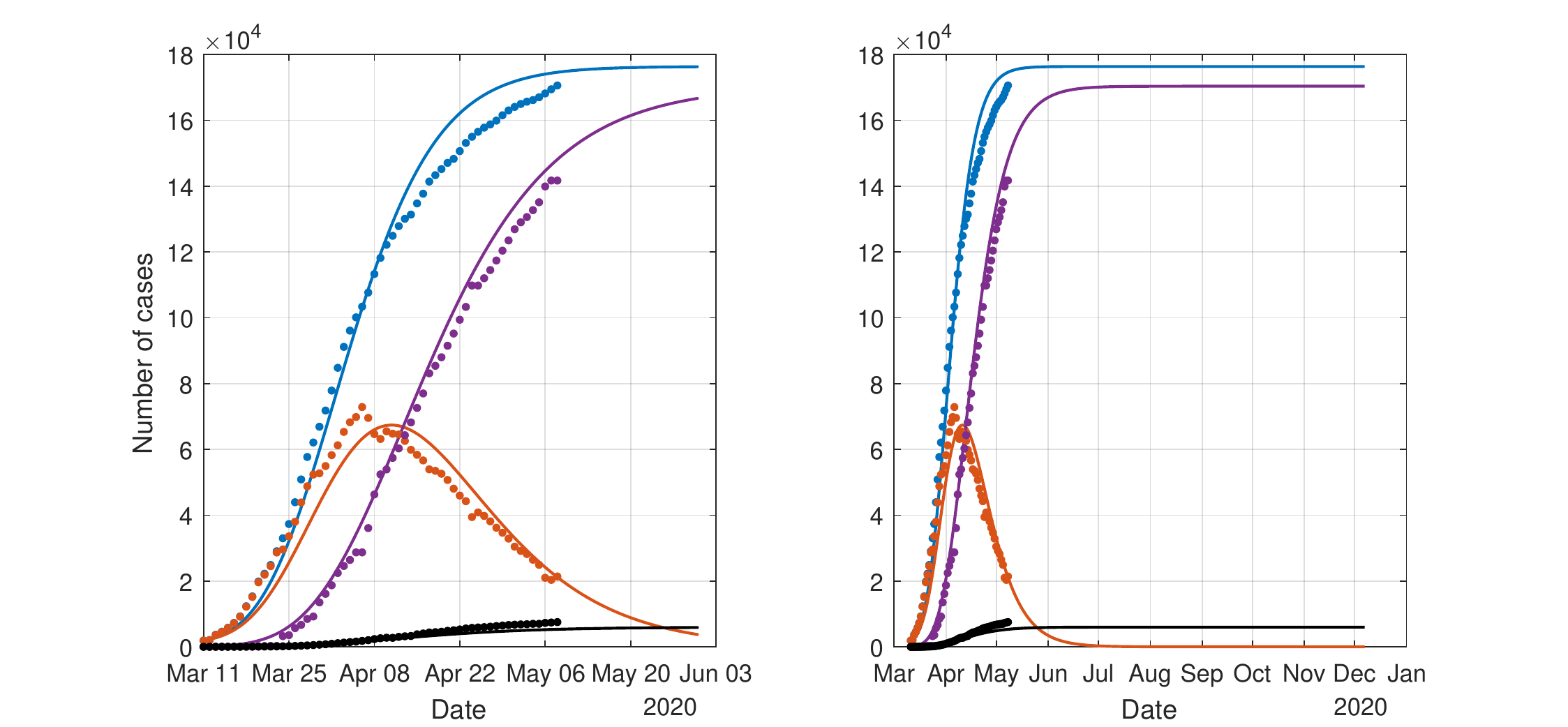}    
\caption{Data fit for Germany using available data between 11 March and 8 May 2020, without a prediction horizon (left) and with a longer prediction horizon to show the final values (right). The figure shows Total cases in blue, Confirmed infectious cases (Q) in orange, Recovered cases (R) in purple, and Deceased cases (D) in black.}
\label{fig:model_fit_Germany}
\end{center}
\end{figure}

\begin{figure}
\begin{center}
\includegraphics[width=12cm]{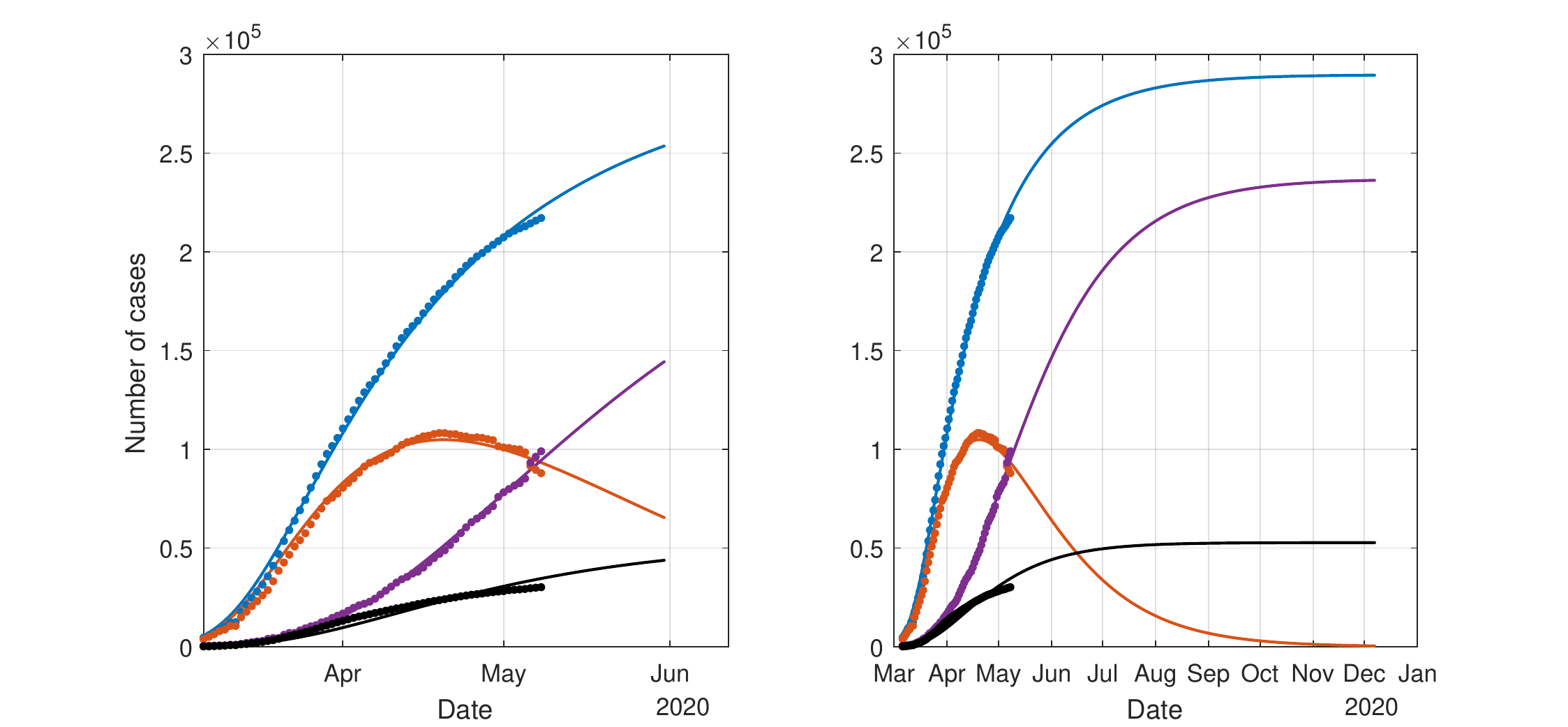}    
\caption{Data fit for Italy using available data between 6 March and 8 May 2020, without a prediction horizon (left) and with a longer prediction horizon to show the final values (right). The figure shows Total cases in blue, Confirmed infectious cases (Q) in orange, Recovered cases (R) in purple, and Deceased cases (D) in black.}
\label{fig:model_fit_Italy}
\end{center}
\end{figure}

\begin{figure}
\begin{center}
\includegraphics[width=12cm]{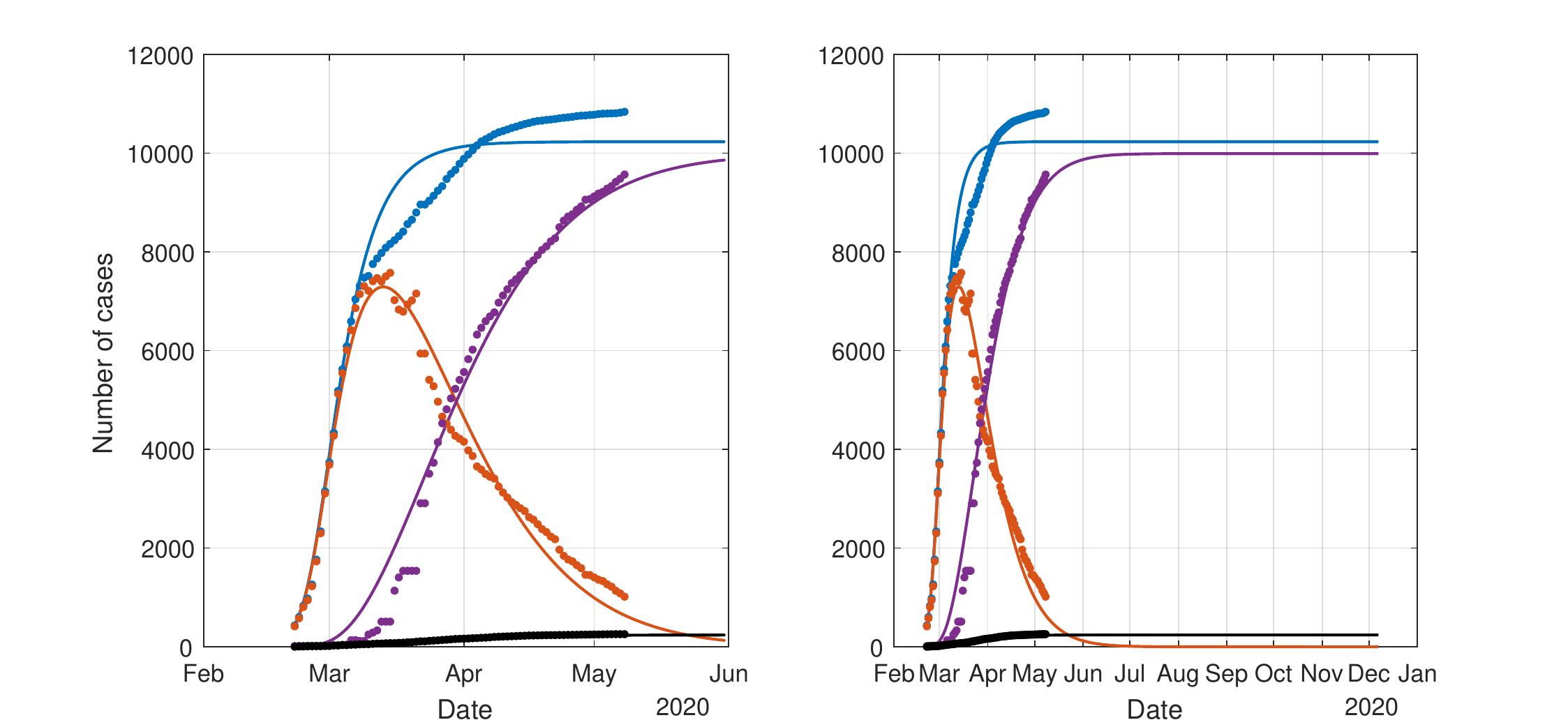}    
\caption{Data fit for South Korea using available data between 22 February and 8 May 2020, without a prediction horizon (left) and with a longer prediction horizon to show the final values (right). The figure shows Total cases in blue, Confirmed infectious cases (Q) in orange, Recovered cases (R) in purple, and Deceased cases (D) in black.}
\label{fig:model_fit_SKorea}
\end{center}
\end{figure}

Germany, Italy, and South Korea have all successfully effected a decrease in the number of daily new cases at the time of writing. This implies enough shape in the curves to obtain a proper fit. The fit of the model to the data for Germany is shown in Fig.~\ref{fig:model_fit_Germany}, the fit for Italy is shown in Fig.~\ref{fig:model_fit_Italy}, and the fit for South Korea is shown in Fig.~\ref{fig:model_fit_SKorea}.

It is interesting to note from Fig.~\ref{fig:model_fit_Germany}, Fig.~\ref{fig:model_fit_Italy}, and Fig.~\ref{fig:model_fit_SKorea} that the mortality rates for all countries are quite different. The progression of the number of cases as well as the fraction of the population that is projected to be infected in total is also quite different across these countries. The SEIQRDP model is however able to fit all the data quite well.

\subsection{Parameter estimation for South Africa}
\label{sec:fitting_RSA}

\begin{figure}
\begin{center}
\includegraphics[width=12cm]{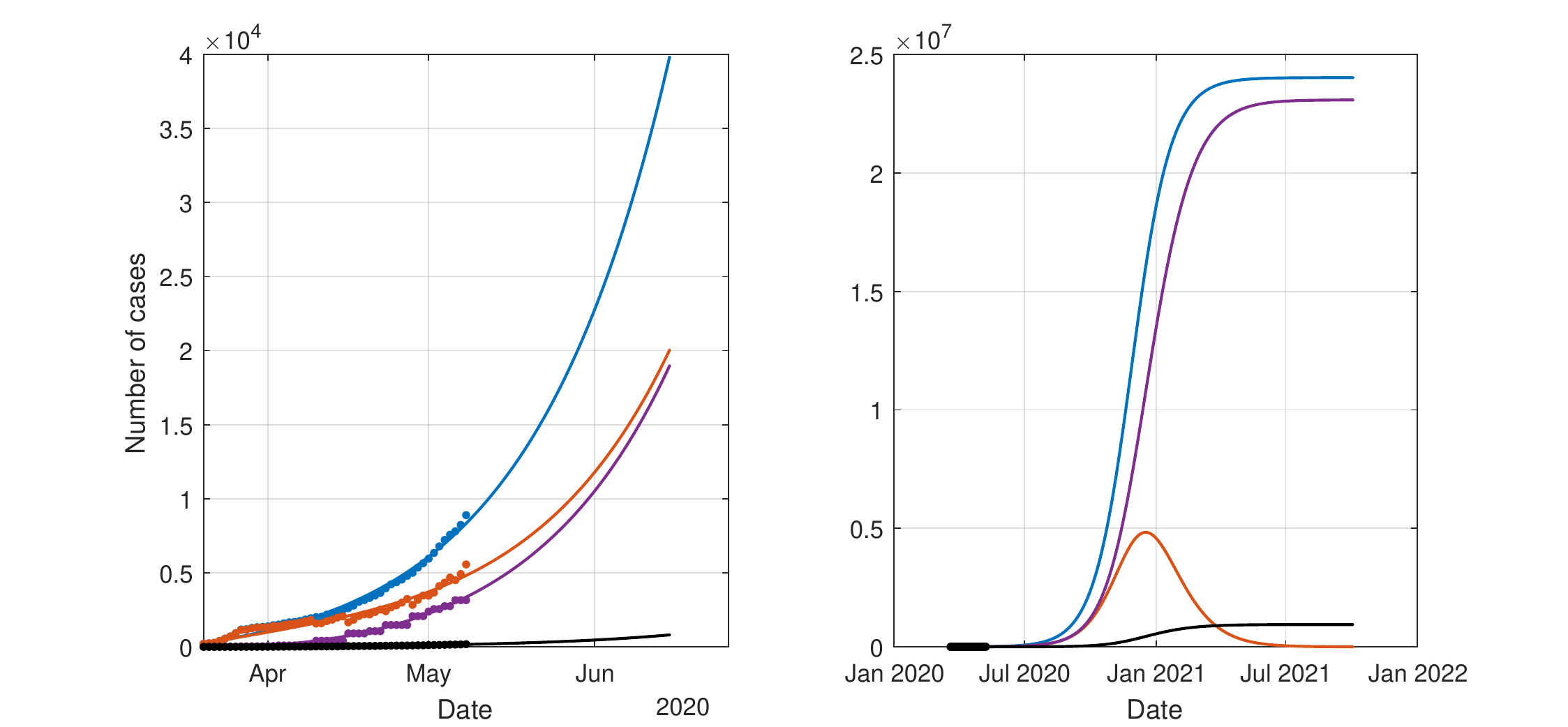}    
\caption{Initial data fit for South Africa where parameter estimation found a constant $\beta(t) = 0.599$ using data from 20 March to 8 May 2020, without a prediction horizon (left) and with a longer prediction horizon (right). The figure shows Total cases in blue, Confirmed infectious cases (Q) in orange, Recovered cases (R) in purple, and Deceased cases (D) in black.} 
\label{fig:model_fit_RSA_initial}
\end{center}
\end{figure}

The number of cases in South Africa only really started to increase in March 2020. As such, the data taken for fitting is only from 23 March to 8 May 2020. The initial result obtained for South Africa is shown in Fig.~\ref{fig:model_fit_RSA_initial} ($\mathrm{MSE = 9.18 \times 10^4}$; $\mathrm{R^2 = 0.965}$).

The $\beta$-value obtained for the model outputs shown in Fig.~\ref{fig:model_fit_RSA_initial} was a constant 0.748, which is much higher than the final $\beta$ values obtained for Germany, Italy, and South Korea. Consequently the projection for the final number of cases is extremely high (24 million). This may be attributed to the fact that South Africa is still in the initial phase of the pandemic, where the number of cases increase rapidly. The predictions obtained from the model during this phase of the pandemic are extremely sensitive to $\beta$, and hence also the basic reproduction number, as given in (\ref{eq:R0}). This fact is illustrated in Section~\ref{sec:sensitivity}.

\begin{figure}
\begin{center}
\includegraphics[width=12cm]{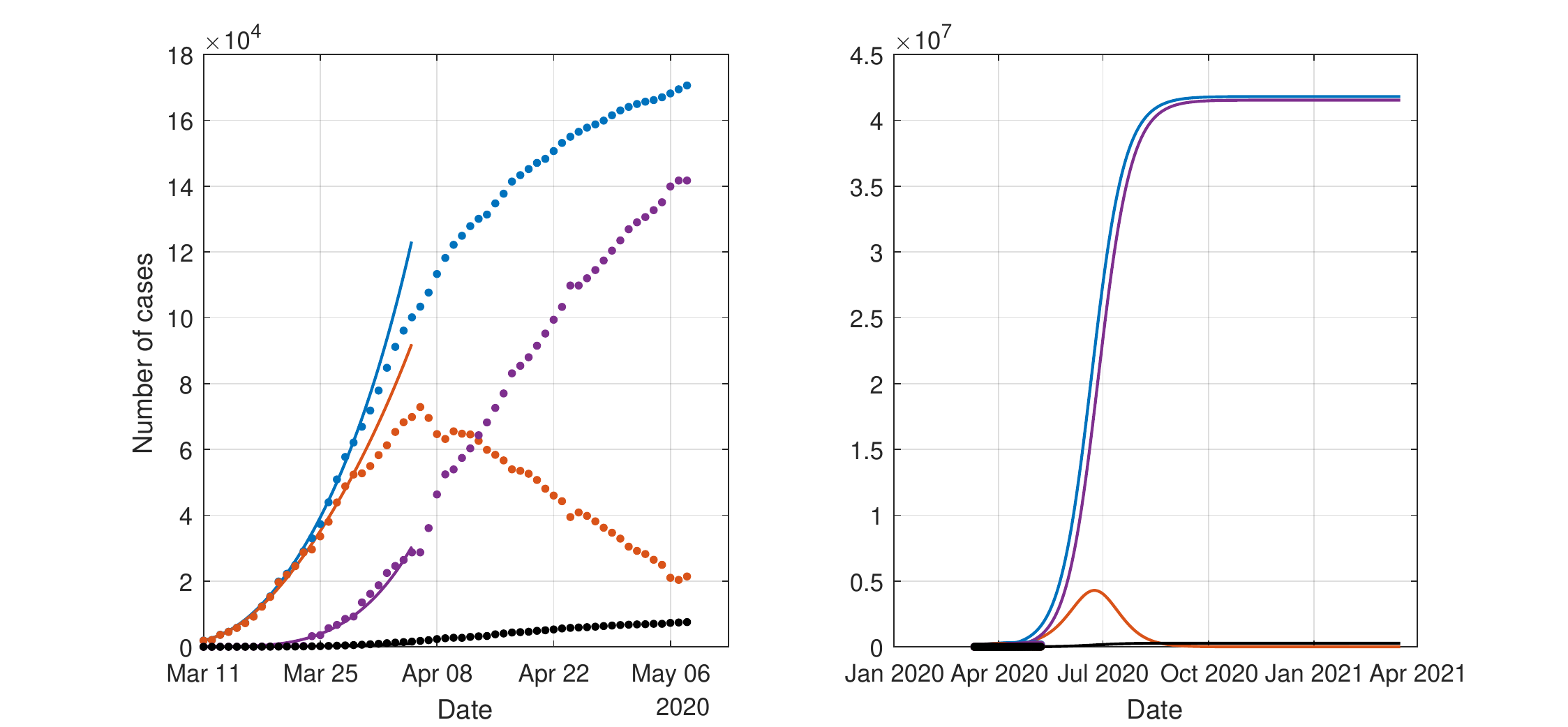}    
\caption{Data fit for Germany only using data from 11 March to 28 March 2020 (i.e. before the number of cases starts to flatten out), without a prediction horizon (left) and with a longer prediction horizon (right). The figure shows Total cases in blue, Confirmed infectious cases (Q) in orange, Recovered cases (R) in purple, and Deceased cases (D) in black.} 
\label{fig:model_fit_GER_early}
\end{center}
\end{figure}

COVID-19 has progressed much further in Germany, Italy, and South Korea than it has in South Africa. One would thus expect that the predictive power of the models obtained for these countries to be better than that of the South African model. The first South African model (with $\beta = 0.599$) of which the outputs are shown in Fig.~\ref{fig:model_fit_RSA_initial}, seems to grossly overestimate the final number of cases. The question therefore arises: what would the predictive power of e.g. the German model have been given the same phase of disease progression than South Africa is now in? To answer this question, the data fit for Germany is repeated, but only data up to the 28th of March are used to determine the model parameters. The result is shown in Fig.~ \ref{fig:model_fit_GER_early}. The model fits the data provided quite well ($\mathrm{MSE = 1.39 \times 10^6}$; $\mathrm{R^2 = 0.994}$), but because the value of $\beta$ cannot be accurately determined, the projection is that by mid-May the number of confirmed cases will be in excess of one million, with the peak in the confirmed cased only occurring by the end of June 2020. In reality, the confirmed cases in Germany appear to flatten out below 20,000 with the peak having occurred in April 2020 (see Fig.~\ref{fig:model_fit_Germany}). This illustrates that it is very difficult to accurately predict the transmission rate parameter ($\beta$) during the initial phase of the pandemic.

\begin{figure}
\begin{center}
\includegraphics[width=12cm]{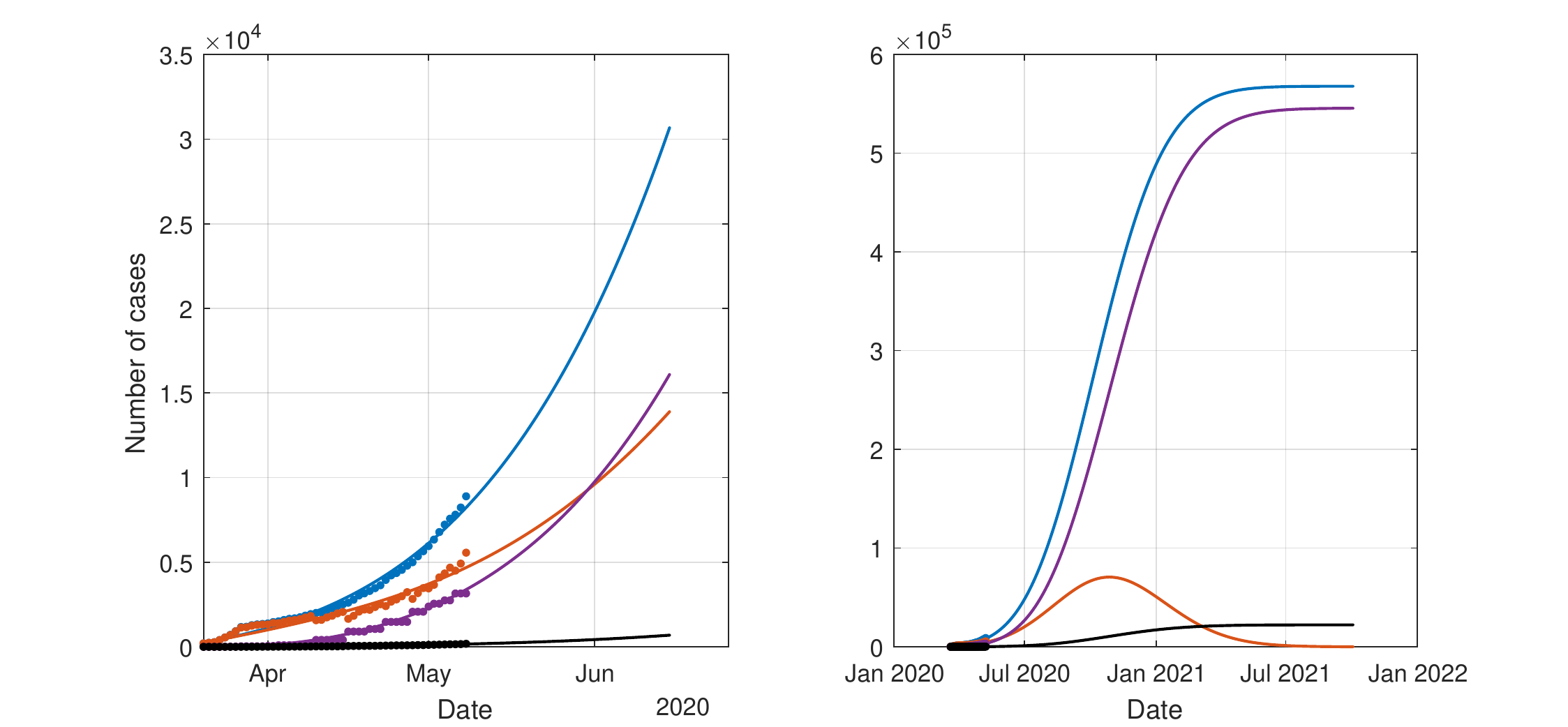}    
\caption{Final data fit for South Africa with $R_0$ constrained using data from 20 March to 8 May 2020, without a prediction horizon (left) and with a longer prediction horizon (right). The figure shows Total cases in blue, Confirmed infectious cases (Q) in orange, Recovered cases (R) in purple, and Deceased cases (D) in black.} 
\label{fig:model_fit_RSA_final}
\end{center}
\end{figure}

To improve the predictive power of the South African model, $\beta$ was therefore constrained to have a final value of 0.25 ($\beta_1 = 0.25$), which is more in line with the final value obtained for e.g. the Italian model. The final fit achieved for the South African model is shown in Fig.~\ref{fig:model_fit_RSA_final} with the parameter values given in Table 1. The South African model with the reduced $\beta$ value still fits the data quite well even though the projections for the final number of cases is much lower, 570,000 as opposed to 24 million, than when $\beta = 0.599$ as shown in Fig.~\ref{fig:model_fit_RSA_initial}.

The final model for South Africa predicts the peak number of infectious cases (about 70,000) to happen in late October (27 October 2020). The total number of cases predicted is about 570,000. These predictions need to be taken in light of the sensitivity of the model to the value of $\beta$, which in turn is also dependent on the number of contacts per person per unit of time. This implies that the number of cases may in future increase or decrease dramatically depending on the regulations in place, and the adherence of the population to those regulations.

\subsection{Parameter estimation results}

The model parameter values obtained are shown in Table~\ref{tb:fitting_parameters}. The total population size ($N$) is simply the total population of the country. The Germany and Italy data were taken from \cite{Eurostat:20} and the South Korea and South Africa data were taken from \cite{WorldBank:20}. These are included in Table~\ref{tb:fitting_parameters} for reference.

One criticism of deterministic epidemiological models (see e.g. \cite{Britton:10}) is that if $R_0 < 1$ there will only be a small outbreak, and if $R_0 > 1$ there will be a major outbreak. This is because the model assumes that the community is homogeneous and that individuals mix uniformly. In reality however individuals will not mix uniformly, especially if regional travel is prohibited. This means that the effective reproduction number does decrease over time in practice. In fact \cite{AlHussein:20} also found a decreasing reproduction number when fitting the same SEIQRDP model to data for Iraq. \cite{AlHussein:20} however achieved this decrease by using $\alpha > 0$.

Given that there is presently no vaccine, having $\alpha > 0$, which allows individuals to move directly from the susceptible portion of the population to the insusceptible portion, seems unlikely. Another possibility is to use an effective population size smaller than the total population of the country like \cite{Fanelli:20}, which also means that the epidemic will abate sooner. \cite{Postnikov:20} notes this situation, where the effective population is much smaller than the actual population, as a ``weak outbreak''.

Both of these options will however not leave room for much of a second peak, which infectious disease experts warn may occur \citep{Prem:20} if regulations are relaxed and the number of contacts per person is allowed to increase significantly. Secondary outbreaks have recently occurred in the Chinese provinces of Jilin and Heilongjiang \citep{Reuters:20}. The option of making $\beta$ a function of time leaves the susceptible population in tact, which might be a more realistic scenario.

\begin{figure}
\begin{center}
\includegraphics[width=12cm]{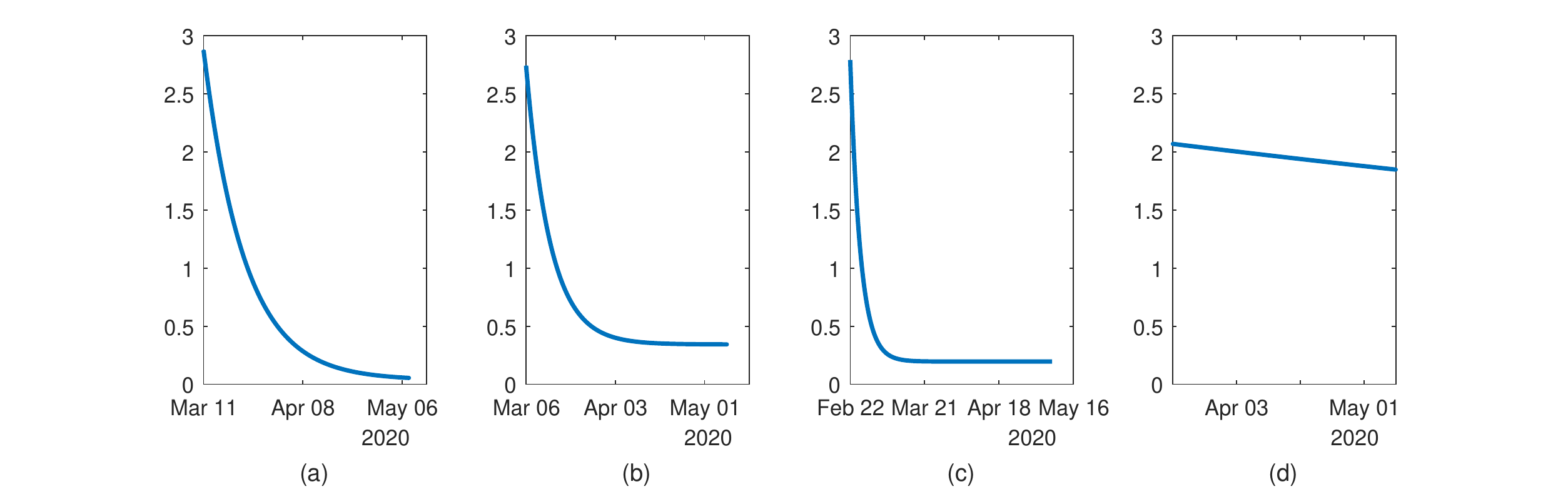}    
\caption{Basic reproduction number ($R_0$) as a function of time for (a) Germany, (b) Italy, (c) South Korea, and (d) South Africa.} 
\label{fig:R0}
\end{center}
\end{figure}

Fig.~\ref{fig:R0} shows the basic reproduction number ($R_0$) as a function of time for all 4 countries modelled. This figure is an indication of the swiftness and strictness of preventative measures, as well as adherence to those measures by the citizens of the country concerned.

\section{Sensitivity analysis}
\label{sec:sensitivity}

\begin{figure}
\begin{center}
\includegraphics[width=12cm]{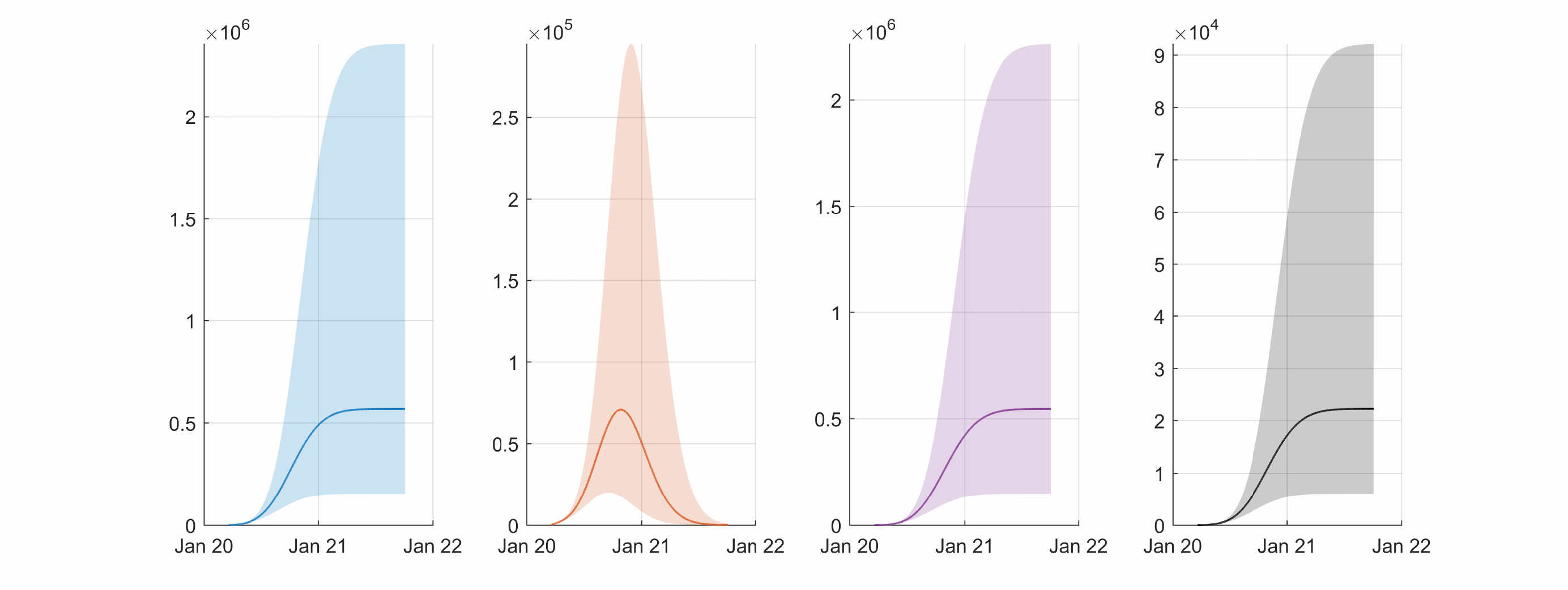}    
\caption{Sensitivity analysis for varying values of $\beta_1$. The solid line shows the nominal value, and the filled patch the range for 10\% smaller and 10\% larger values. Total cases are shown in blue, Confirmed infectious cases in orange, Recovered cases in purple, and Deceased cases in black.}
\label{fig:Sensitivity_beta}
\end{center}
\end{figure}

\begin{figure}
\begin{center}
\includegraphics[width=12cm]{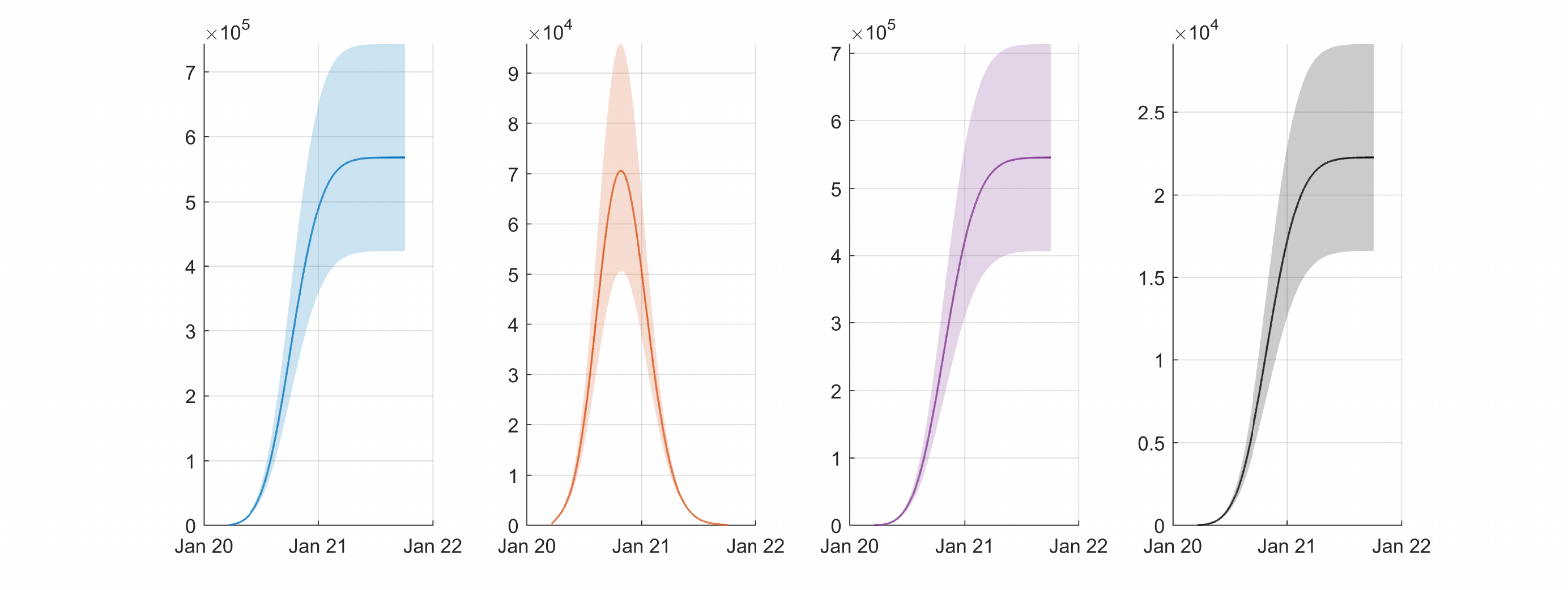}    
\caption{Sensitivity analysis for varying values of $\gamma$. The solid line shows the nominal value, and the filled patch the range for 10\% smaller and 10\% larger values. Total cases are shown in blue, Confirmed infectious cases (Q) in orange, Recovered cases (R) in purple, and Deceased cases (D) in black.}
\label{fig:Sensitivity_gamma}
\end{center}
\end{figure}

\begin{figure}
\begin{center}
\includegraphics[width=12cm]{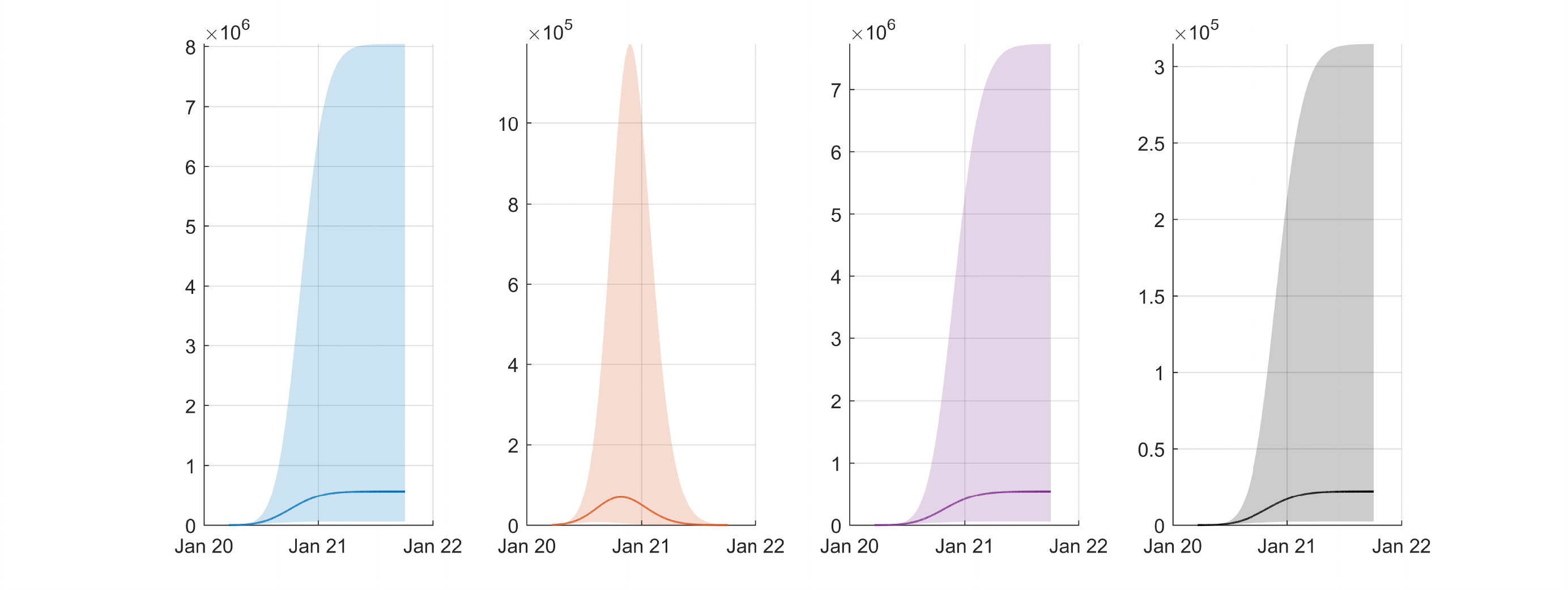}    
\caption{Sensitivity analysis for varying values of $\delta$. The solid line shows the nominal value, and the filled patch the range for 10\% smaller and 10\% larger values. Total cases are shown in blue, Confirmed infectious cases (Q) in orange, Recovered cases (R) in purple, and Deceased cases (D) in black.}
\label{fig:Sensitivity_delta}
\end{center}
\end{figure}

\begin{figure}
\begin{center}
\includegraphics[width=12cm]{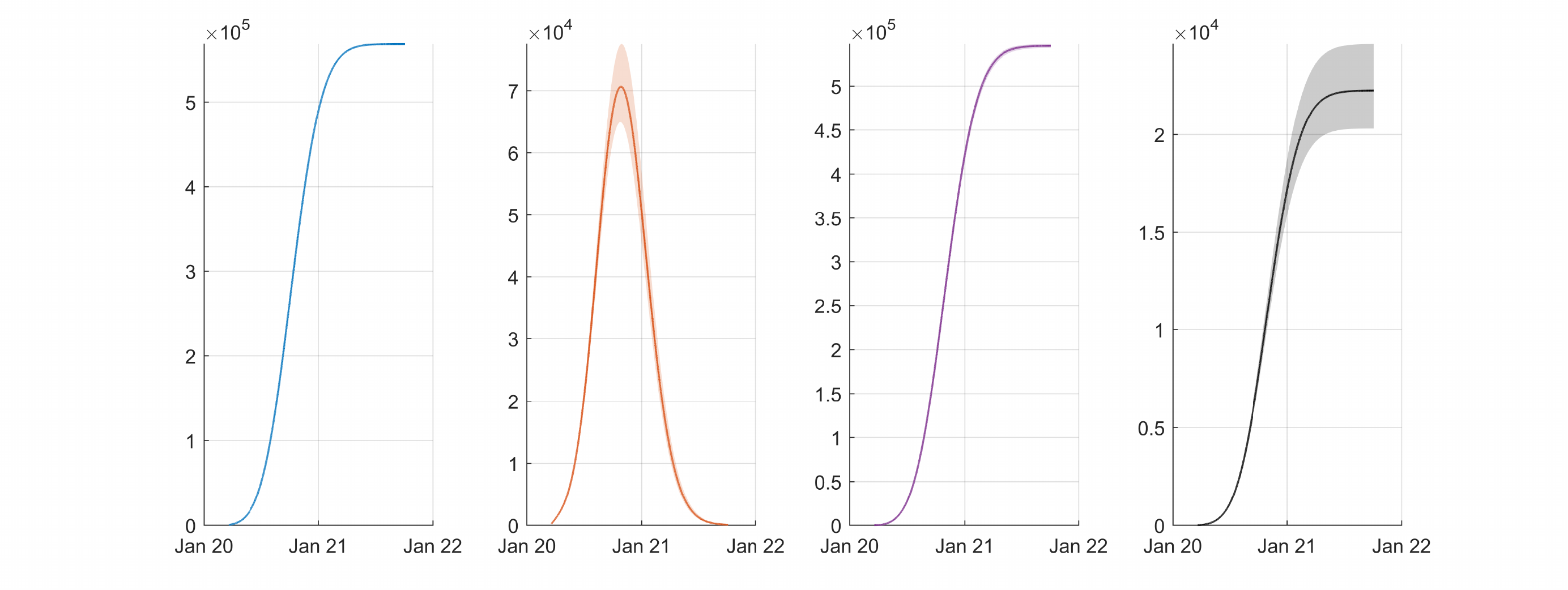}    
\caption{Sensitivity analysis for varying values of $\lambda_1$. The solid line shows the nominal value, and the filled patch the range for 10\% smaller and 10\% larger values. Total cases are shown in blue, Confirmed infectious cases (Q) in orange, Recovered cases (R) in purple, and Deceased cases (D) in black.}
\label{fig:Sensitivity_lambda}
\end{center}
\end{figure}

\begin{figure}
\begin{center}
\includegraphics[width=12cm]{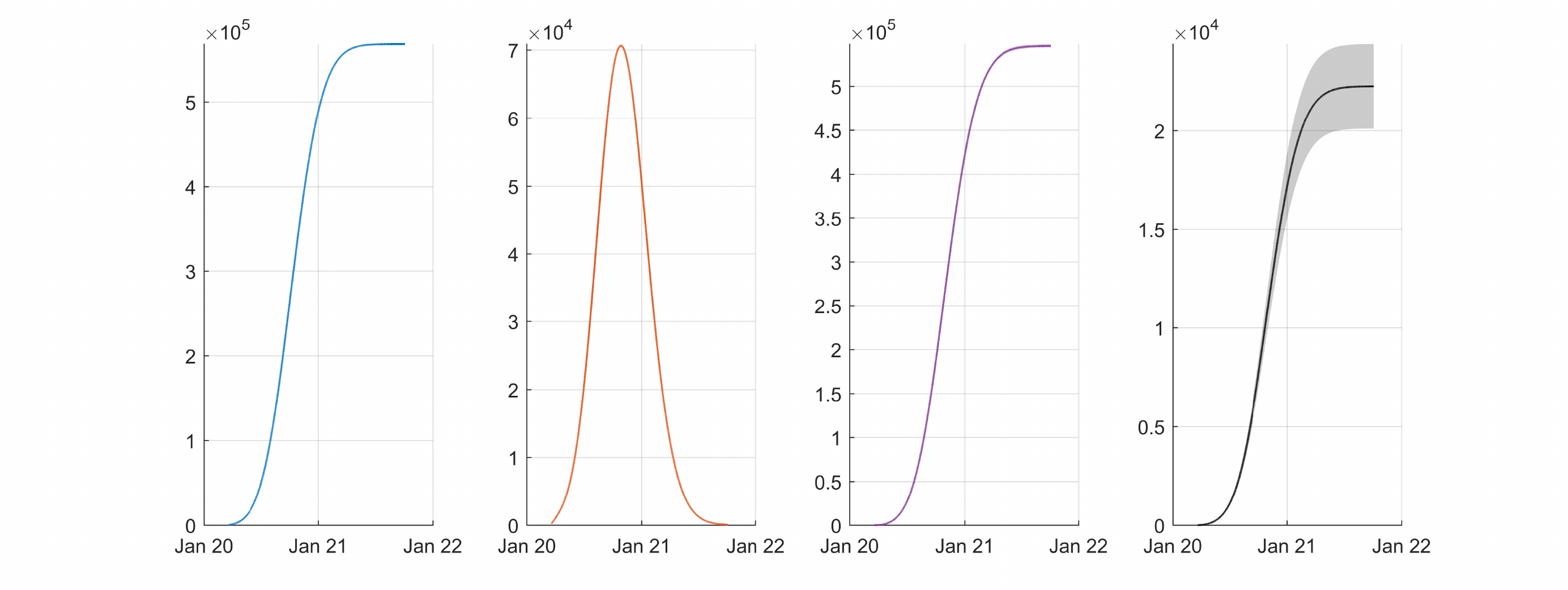}    
\caption{Sensitivity analysis for varying values of $\kappa_1$. The solid line shows the nominal value, and the filled patch the range for 10\% smaller and 10\% larger values. Total cases are shown in blue, Confirmed infectious cases (Q) in orange, Recovered cases (R) in purple, and Deceased cases (D) in black.}
\label{fig:Sensitivity_kappa}
\end{center}
\end{figure}

The predictive power of the models obtained in Section~\ref{sec:parameter_estimation} rely heavily of the accuracy of the parameters obtained. This is especially true in the South African case where COVID-19 has not progressed far enough for the confirmed infectious cases to have peaked. In order to illustrate the sensitivity of the South African model to the parameters, the base parameters are decreased by 10 \% and then increased by 10 \%, one at a time, and the simulation over time repeated. This is done for $\beta_1$, $\gamma$, $\delta$, $\lambda_1$, and $\kappa_1$. The results are shown in Fig.~\ref{fig:Sensitivity_beta} - \ref{fig:Sensitivity_kappa}.

Fig.~\ref{fig:Sensitivity_beta} shows the impact that varying $\beta_1$ has on the total number of cases. It confirms that a small change in the transmission rate over the initial period when the number of cases increases exponentially has a very large effect. This also indicates that reducing the possibility of transmission early on should be effective in limiting the total number of cases.

From Fig.~\ref{fig:Sensitivity_gamma} it is clear that changing $\gamma$ has an impact on the number of cases, but the impact is much smaller than that of changing $\beta_1$. $\gamma$ drives the movement of cases from the ``Exposed'' to the ``Infectious'' compartment; i.e. the time between being exposed to being infectious. If this period is longer there are slightly fewer cases in the ``Infectious'' compartment, and hence slightly fewer people that can infect others.

It is clear from Fig.~\ref{fig:Sensitivity_delta} that the effect of $\delta$ on the number of cases is also very pronounced. Mathematically this makes sense as it can be seen from equation (\ref{eq:R0}) that changes in $\delta$ have similar effects as changes in $\beta_1$ as they affect the basic reproduction number in a similar way. Practically it also makes sense that the longer a person remains in the ``Infectious'' category (with the potential to infect others), without having been confirmed infectious such that they can isolate (i.e. move into the ``Quarantined'' category), the number of new cases generated will increase.

This is also where testing plays an important role. The more tests that are conducted, the sooner ``Infectious'' cases will be identified. This means $\delta$ increases and the total number of cases generated decreases.

Changing $\lambda_1$ or $\kappa_1$ does not have any impact on the total number of cases (as can be seen from Fig.~\ref{fig:Sensitivity_lambda} and \ref{fig:Sensitivity_kappa}). The only real effect of these parameters is whether cases will end up in the ``Recovered'' or ``Deceased'' compartments. On the scale of the number of ``Recovered'' cases shown in the figures, the effect is almost not visible.

\section{Conclusion}
\label{sec:conclusion}

An epidemiological model was developed for spread of COVID-19 in South Africa. A variant of the classical compartmental SEIR model, called the SEIQRDP model, was used. The SEIQRDP model was first parameterized on data for Germany, Italy, and South Korea, to test the efficacy of the model. Good fits were achieved with reasonable predictions of where the number of COVID-19 confirmed cases, deaths, and recovered cases will end up and by when. 

South African data for the period from 23 March to 8 May 2020 was then used to obtain SEIQRDP model parameters. It was found that the model fits the initial disease progression well, but that the long-term predictive capability of the model was rather poor. The South African SEIQRDP model was subsequently recalculated with the basic reproduction number constrained to reported values. The resulting model fit the data well, and long-term predictions appear to be reasonable. The peak number of the confirmed infectious individuals is predicted to happen at the end of October 2020, and the total number of deaths is predicted to range from about 10,000 to 90,000, with a nominal value of about 22,000. All of these predictions are heavily dependent on the disease control measures in place, and the adherence of the population to these measures. The study concludes with a sensitivity analysis that show that the model is particularly sensitive to parameters used to determine the basic reproduction number. 

The aim is to in future use a feedback control approach together with the South African SEIQRDP model to determine the epidemiological impact of varying lockdown levels proposed by the South African Government.

\raggedright
\bibliography{ifacconf}             

\end{document}